\documentclass[twocolumn]{jpsj2} 
\title{
Ultrasound Study of the Solid-Liquid Transition and Solid-Liquid Interface of $^4$He in Aerogels 
}

\author{Koichi \textsc{Matsumoto}\thanks{E-mail address: kmatsu@kenroku.kanazawa-u.ac.jp}, 
Hiroyuki \textsc{Tsuboya}\thanks{Kyowa Electronic Instruments Co., Ltd. 3-5-1, Chofugaoka, Chofu, Tokyo 182-8520, Japan}, 
Keiichi \textsc{Yoshino}\thanks{Nippon Electric Glass Co., Ltd. 7-1, Seiran 2-chome, Otsu, Shiga 520-8639, Japan}, 
Satoshi \textsc{Abe}, 
Hiroyuki \textsc{Tsujii}, and 
Haruhiko~\textsc{Suzuki}}

\inst{Department of Physics, Kanazawa University, Kanazawa 920-1192, Japan \\
}

\abst{
The freezing and melting of $^{4}$He in various aerogels with porosities ranging from 92 to 97\% were studied using longitudinal ultrasound.
The freezing pressure, detected from changes in velocity and attenuation, was elevated compared with that of the bulk by about 0.3~MPa, and showed a weak dependence on aerogel porosity. 
Inside the aerogels, there was a liquid phase, a solid phase or a coexisting state depending on temperature and pressure of the sample. 
We report the measurement of the transmission of sound through the solid-liquid interface in aerogel for the first time, 
which was independent of temperature and unaffected by a small addition of $^3$He, unlike that at the bulk interface. 
This indicates that sound attenuation at the interface is due to the disorder, originating from silica strands. 
Small amounts of $^3$He in the solid significantly decreased attenuation, 
because the resultant pinning of dislocations by $^3$He suppressed phonon scattering, as is observed in bulk solid $^4$He. 
}

\kword{$^4$He, crystallization, aerogel, ultrasound, porous media.}

\begin{document}
\maketitle

\section{Introduction} 

$^4$He confined in porous media is an interesting interacting Bose system of nanometer scale, and has been extensively studied, experimentally and theoretically, since interatomic interactions can be controlled by adjusting pore structure, size, and $^4$He density. 
It has been shown that the pressure-temperature ($p$-$T$) phase diagram is significantly altered for helium in porous materials. 
Experiments on the superfluid transition of $^4$He contained in porous media, such as aerogel, xerogel and Vycor glass, have revealed that the superfluid transition temperature is lower than that of bulk $^4$He \cite{Wong,Chan}. 
As for the solid-liquid transition, a number of works\cite{AdamsTaungUhig, BeamishHikata, BeamishSolid, YamamotoShirahama, MolzBeamish} reported freezing-pressure elevation in various porous materials, which is ascribed to the inhibition of solid nucleation in narrow pores.

There are common features of solid-liquid transitions in a number of porous materials. 
The freezing and melting transitions broaden; however, an abrupt onset of freezing and the completion of melting indicate a sharp upper cutoff in pore size distribution. 
Transition temperature decreases in inverse proportion to the pore size from that of the bulk, and show pronounced hysteresis. 
Simple supercooling at the first-order phase transition cannot elucidate such hysteresis, since melting occurs at a lower temperature than that in the bulk.

The $p$-$T$ phase diagram has been well established for Vycor glass, which has three-dimensionally connected random pores, 6 nm in diameter. 
For example, $^4$He in Vycor glass has been found to remain liquid up to 4.2~MPa, and the superfluid transition shifts by about 0.2~K from that of bulk $^4$He\cite{AdamsTaungUhig}. 
Several sound experiments have been performed on $^4$He in Vycor glass\cite{BeamishHikata,WarnerBeamish,BeamishElbaum}. 
Ultrasound measurements are suitable for the study of superfluidity and freezing-melting in porous media because sound velocity and attenuation are sensitive to changes in both the density and elastic modulus of the material in pores. 
In addition, superfluid transition is accompanied by a critical attenuation peak. 
When liquid helium in pores freezes, the solid-liquid interface provides an additional attenuation and a significant change in the velocity of sound, because sound travels about 60\% faster in solid helium than in bulk liquid helium\cite{Greywall, Donnelly}.

It should be interesting to study superfluidity and solid-liquid transition in aerogel because it has a qualitatively different pore structure from other porous materials. 
Silica aerogel can have porosity as large as 99.8\%, and consists of a disordered silica network with fractal mass correlation. 
The aerogel pore structure varies with porosity and type of fabrication method. 
The present work is motivated by questions of how the porosity, structure, and possible fractal nature of aerogels influence the superfluid and solid-liquid transition of $^4$He.

The freezing of $^4$He in aerogel was first reported by Molz and Beamish\cite{MolzBeamish}. 
In their longitudinal ultrasound measurements at 9~MHz, the freezing pressure in 87\% porous aerogel was elevated by about 0.46~MPa at low temperatures, and was lower than that in Vycor glass. 
The hysteresis on freezing and melting was shown to be smaller than that observed in Vycor glass.

Previous studies of the acoustic properties of solid helium, where thermally activated vacancies\cite{LenguaGoodkind} and dislocation motion\cite{GranatoLucke} are thought to play important roles in determining acoustic properties have been reported\cite{Crepeau, Greywall, Goodkind, BurnsGoodkind, LenguaGoodkind, HoGoodkind, HSuzuki1, HSuzuki2}. 
Solid helium in an aerogel should have many dislocations caused by silica strands. 
Recently, Kim and Chan have reported the possible observation of a supersolid helium phase in Vycor glass by torsional oscillator measurements\cite{Kim}. 
It is interesting to study how vacancies and dislocations behave in a disordered helium crystal, because dislocations play an important role in the nonclassical rotational inertia of solid $^4$He\cite{DayBeamishSuperSolid}. 
Our study will shed light on this problem.

The interface between superfluid and solid $^4$He is an ideal system for studying crystal growth, since $^4$He has a high transport of mass and latent heat to the interface during crystallization, because of its superfluidity. 
The latent heat of solidification is nearly zero, as can be seen from $dp/dT\sim 0$ in the $p$-$T$ phase diagram. 
The $^4$He interface also gives us an opportunity to study surface physics. 
Ultrasound transmission through the solid-liquid interface has been studied both experimentally\cite{Castaing, Moelter, Poitrenaud, Amrit} and theoretically in bulk $^4$He\cite{Puech, Nozieres}. 
It has been shown that sound transmission through the interface is significantly affected by crystal growth, which is induced by a difference in chemical potential resulting from acoustic pressure. 
Atomic roughness at the interface plays an important role in crystal growth, and it is interesting to note how disorder at the interface, introduced by the aerogel, influences the sound transmission and crystal growth of $^4$He.

$^3$He atoms, doped into the solid and interface of bulk $^4$He, have a significant effect on sound propagation. The motion of dislocations is pinned by 1\% $^3$He doping into $^4$He crystals, and sound attenuation is suppressed\cite{HSuzuki1}. 
For solid helium in aerogel, a similar pinning mechanism is expected. 
It is known that crystal growth in the bulk is largely influenced by $^3$He impurity\cite{Wang, MSuzuki} because $^3$He selectively remains at the interface, impeding mass and thermal transports across the interface.

In this work, we report ultrasound measurements of $^4$He in various aerogels, with porosities ranging from 92 to 97\%. 
The $p$-$T$ phase diagram has been explored by the freezing and melting of solid $^4$He. 
Sound propagation in the liquid phase has been reported elsewhere\cite{MatsumotoJETP, MatsumotoISSP1, NishikawaMatsumotoISSP2, MatsumotoJLTP}. 
The solid-liquid interface of $^4$He in aerogel has been realized for the first time, and we present details of sound propagation, both in solid and through the interface. 
Finally, the effect of $^3$He impurity is reported, both in solid and at the interface. 
Crystal growth on the disordered interface is also discussed.

\section{Experimental Procedure}

Aerogel samples are prepared by a sol-gel process, followed by hypercritical drying. 
Silica aerogel is thought to be a network of nanoscale SiO$_2$ strands, with a fractal structure over a wide range of scale. 
We performed sound measurements with three aerogels of 92.6, 94.0, and 97.0\% porosity. 
The 97.0\% porous aerogel was manufactured by Matsushita Electric Works, and the others were from AIST Nagoya. 
Similar gels show a wide range of pore sizes, from a few nanometers to several hundreds of nanometers. 
An aerogel with 94.0\% porosity made in the same lot had a typical pore size of about 15~nm, determined from nitrogen adsorption measurements\cite{Tajiri}. 
We have not determined the pore size distribution for all the samples. 
When the pores are filled with liquid helium, normal fluid component clamps to the material owing to viscosity. 
In our ultrasound experiment, the viscous penetration depth of liquid $^4$He is estimated to be larger than the mean silica strand distance. 
The superfluid component is not directly associated with aerogel motion.

Samples were cut into cylinders, and LiNbO$_3$ piezoelectric transducers were mechanically pressed to the sample ends. 
No gluing was used in order to avoid possible damage to the aerogels by the adhesive solvents. 
The side of the aerogel cylinder was connected to the surrounding bulk helium reservoir. 
The 92.6 and 94.0\% porous samples had a 7~mm diameter and a 3.0~mm length, and the 97.0\% porous sample had a 2.3~mm length.  
To measure the frequency dependence, we used two sets of longitudinal transducers with fundamental frequencies of 6 and 10~MHz, since no harmonics could be observed owing to the large attenuation. 

The acoustic cavity was installed in a copper cell, and mounted on a $^3$He cryostat in the direction of vertical sound propagation. 
Temperature was measured with a Cernox (LakeShore) sensor and controlled using a PID controller. 
The pressure of liquid helium was measured with a sensor at room temperature and stabilized using another PID controller. 
Solid helium was obtained by the capillary block method, so that the pressure sensor was unable to measure solid pressure. 
When the fill line was blocked, the pressure at the sample cell was determined using the sound velocity at the completion of freezing of the surrounding bulk $^4$He. 
The detailed pressure determination procedure is described in the following section.

Ultrasonic measurement was conducted using a standard pulse transmission and phase-sensitive detection technique\cite{MatsumotoJETP, MatsumotoISSP1}. 
We used drive amplitudes of less than $-$5~dBm, which were considerably smaller than those used by Iwasa and coworkers\cite{HSuzuki1, HSuzuki2}, and comparable to those used by Goodkind\cite{Goodkind}. 
All sound measurements were taken in the linear region, where the amplitude of the received signal was proportional to input signal amplitude. 
In order to improve signal-to-noise ratio, several hundreds of sound signals were averaged after temperature stabilization. 
The resolution of the phase detection was within 10~ppm for changes in sound velocity. 
Absolute sound velocity was determined using the flight time of the sound signal to within about 2\%.

\section{Results and Discussion}

\subsection{Freezing of $^4$He in aerogel}

\begin{figure}[tb]
\begin{center}
\includegraphics[width=80mm]{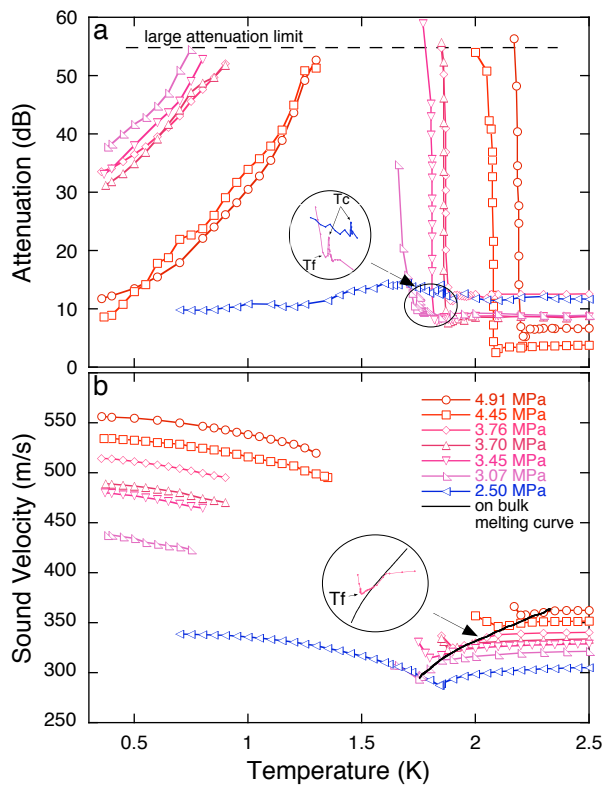}
\end{center}
\caption{(Color online) Freezing of $^4$He in 97\% porous aerogel. 
Attenuation and velocity of 6~MHz longitudinal ultrasound are shown in (a) and (b), respectively; the same measurements of liquid phase at 2.5~MPa are also shown for comparison. 
The freezing pressure $p_f$ is listed in the legend. 
The inset of (a) shows critical attenuation, ascribed to the superfluid transition under low-pressure conditions. 
The solid line without any symbol in (b) represents velocity of liquid on the bulk melting curve. 
The inset of (b) is the expanded sound velocity change around the freezing onset at 3.76 MPa. 
}
\label{970-6MHz_AttC}
\end{figure}

Figures~\ref{970-6MHz_AttC}(a) and \ref{970-6MHz_AttC}(b) show temperature variation of the attenuation and velocity for 6~MHz ultrasound in the 97.0\% porous aerogel for seven cooling processes. 
The sound velocity on the bulk melting curve is shown as a solid line in Fig.~\ref{970-6MHz_AttC}(b). 
The freezing pressure of $^4$He in porous media is elevated from that of the bulk. 
The cell was pressurized at high temperature and cooled, while cell pressure was kept constant from 2.5 K to the melting curve of bulk solid. 
The temperature was carefully kept at the bulk melting point to grow bulk solid outside of the aerogel. 
After the fill line became blocked, the direct monitoring of liquid pressure in aerogel was impossible and one had to use velocity to deduce internal pressure.
On further cooling, the pressure in the cell began to decrease because the quantity of $^4$He remained constant, since the fill line was blocked. 
The temperature dependence of sound velocity of $^4$He in aerogel follows that of the liquid along the bulk melting curve, as shown by the solid line in Fig.~\ref{970-6MHz_AttC}(b). 
After the complete freezing of the surrounding bulk $^4$He at a pressure $p_f$, the liquid in the aerogel was cooled to the freezing onset with little pressure variation because of the small thermal expansion of liquid $^4$He. 
Freezing pressure was determined as $p_f$ similarly to that in other studies\cite{BeamishSolid, MolzBeamish}. 
The velocity of sound becomes higher than it is at the bulk melting pressure, indicating that liquid $^4$He in aerogel has a higher pressure than that on the bulk melting curve. 
The freezing temperature of $^4$He in aerogel, $T_f$, was determined by locating abrupt increases in sound attenuation and velocity as shown in Fig.~\ref{970-6MHz_AttC}.

Using the pressure dependence of sound velocity in the liquid phase measured at $T_f$, the pressure at $T_f$ was evaluated by extrapolating velocity to the freezing onset. 
The estimated pressure agrees with $p_f$ because of the small density change with temperature. 
Hereafter we use $p_f$ in indicating the freezing pressure in aerogel, as well. 

The onset of freezing was very sharp, and independent of ultrasound frequency. 
When the cell was cooled further at $p_{f}=$ 3.07~MPa, the attenuation peak and minimum sound velocity, which correspond to the superfluid transition, were observed at $T_{c}$, which was slightly lower than that observed at melting pressure. 
This shows that $^4$He froze from the superfluid phase at 3.07~MPa; however, we saw no superfluid transition above 3.4~MPa, indicating that freezing in the aerogel occurred from the normal phase. 

The usual interpretation of freezing is based on the nucleation theory, where crystal growth begins with microscopic nucleation. 
Sasaki et al.\cite{SasakiJLTP} have optically observed solid $^4$He grown by the blocked capillary method in bulk helium. 
Because a cell cools from its sides, crystallization in the bulk starts on the wall of the cell. 
Nucleation in aerogels occurs in the vicinity of piezotransducers, since the cell is slightly warmer at the center owing to the low thermal conductivity of liquid in the normal phase and aerogel. 
During cooling, crystal growth in aerogel would proceed towards the center as in the bulk. 
The highly polycrystal solid $^4$He is observed in aerogel\cite{MuldersPRL08}. 
Sound signals disappeared after the onset of freezing owing to the large attenuation of the solid-liquid interface and/or solid. 
Attenuation decreased with decreasing temperature, and the signal could be observed again in lower temperature. 
Sound velocity increased with $p_f$ at the lowest temperature. 
On the other hand, the attenuation was large and almost independent of $p_f$ between 3.07 and 3.76~MPa, and became smaller above 4.45~MPa. 
The velocity and attenuation behaviors suggest that $^4$He was partially frozen under low-pressure conditions.

At $p_f$ values of 4.91 and 4.45~MPa, the sound velocity was about 550~m/s, close to that of bulk solid even though sound velocity depends on crystal orientation. 
Considering the high velocity and small attenuation, $^4$He in the acoustic cavity was completely frozen somewhere between 3.76 and 4.45~MPa. 
On the other hand, at low $p_f$, sound velocity varied from that in liquid to solid and had a nearly linear dependence on $p_f$, as shown in Fig.~\ref{970-6MHz_AttC}(b). 
The solid-to-liquid fraction increased with increasing $p_f$ and that obtained from the sound velocity varied from 51 to 85\% with increasing $p_f$ from 3.07 to 3.76 MPa. 
When the molar volume change in aerogel is assumed to be about half of that in the bulk as reported in other porous materials\cite{AdamsTaungUhig, BittnerAdams}, the solid-to-liquid fraction obtained from $T_f$ and $p_f$ is in reasonable agreement with that obtained from sound velocity.

Attenuations in liquid and solid are almost the same at the lowest temperature. 
An increase from those in liquid and solid is regarded as the attenuation caused by the solid-liquid interface. 
Attenuation due to the interface is insensitive to the solid-to-liquid fraction. 
These behaviors of velocity and attenuation support the existence of a macroscopic interface. 
In the solid-liquid coexisting case in the bulk, the liquid region moved up due to the buoyancy force\cite{SasakiJLTP}. 
If the buoyancy force is strong enough to move liquid in aerogels, a single solid-liquid interface may also occur. 
However, if the interfacial tension between aerogel and helium prevents liquid from moving up, the liquid phase would be located at the center. Considering the small active area of piezotransducers, the sound path will be solid-liquid-solid.

\subsection{Hysteresis}

A small hysteresis of freezing and melting was observed in our aerogels. 
The differences between freezing $T_f$ and melting $T_m$ were between 40 and 90 mK for the three aerogels studied. 
We can draw information about crystal nucleation from the hysteresis curve. 
In Vycor glass, it is well known that the melting temperature is not equal to the freezing temperature, but the hysteresis in the aerogel is much smaller than those observed in Vycor glass and Bioglass\cite{BittnerAdams}. 
The completion of melting occurred at a lower temperature than that of bulk melting, indicating that hysteresis is not merely due to a supercooling effect. 
The pore distribution in the aerogels seems to be wider than that in Vycor glass, and the small hysteresis may result from the more open structure of the aerogels. 

\begin{figure}[tb]
\begin{center}
\includegraphics[width=80mm]{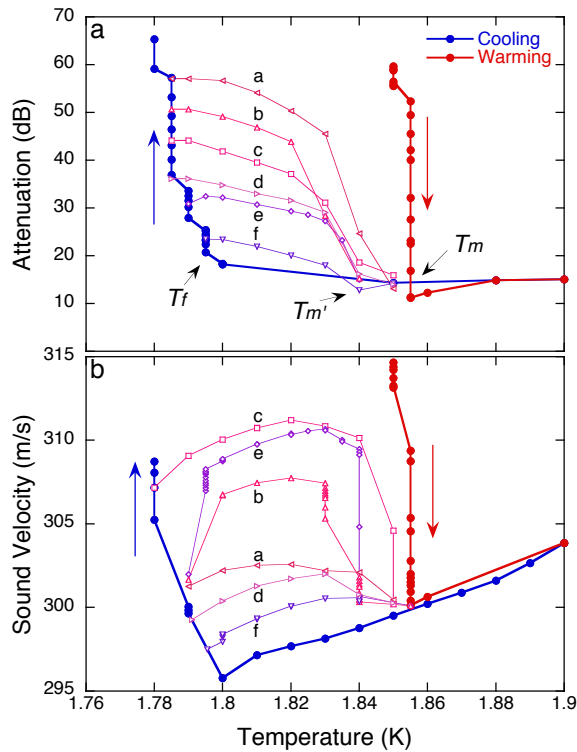}
\end{center}
\caption{(Color online) Attenuation and velocity of $^4$He in 97.0\% porous aerogel. 
The curves labeled as a-f are hysteresis loops with the lowest temperatures successively approaching the onset of freezing at $T_{f}=$1.795~K. 
}
\label{Hysteresis}
\end{figure}

A number of measurements were made, in which cooling was stopped at different temperatures below the onset of freezing. 
Figures~\ref{Hysteresis}(a) and \ref{Hysteresis}(b) show the attenuations and velocities for various cooling and warming processes, at $p_{f}=$3.4~MPa and a frequency of 10~MHz. 
Solid symbols represent those in cooling down and warming up from the lowest temperatures. 
Open symbols represent those obtained when the cell was warmed back to melting temperature from the temperature, which is slightly below $T_f$. 
The temperature $T_m$ at which $^4$He completed melting is identified as the point at which the warming curve intersects with the cooling curve. 
We cooled the sample from a temperature above $T_m$, where there was no solid in the aerogel, to some temperature below the freezing onset $T_f$, stopped cooling, and then warmed the sample again to a temperature above the complete melting temperature $T_m$. 
The paths in Fig.~\ref{Hysteresis}, which are labeled as a-f, corresponded to the cycles where the lowest temperature reached was incrementally closer to the freezing onset. 

A solid is expected to be nucleated near the transducers as mentioned in the former section. 
The microscopic crystals grow individually at first, resulting in several small crystals distributed near the transducers. 
Sound velocity and attenuation are determined by the distribution of microscopic crystals, which would differ from run to run in our experiments. 
The $^4$He crystal induces attenuation owing to the scattering of sound waves, but sound velocity is not affected significantly until the whole active area of transducers are covered with solid. 
Small crystals eventually connect together as they grow, and this connection decreases surface-to-volume ratio, reducing interface energy. 
The rearrangement of a disordered surface layer, which is believed to exist, may further reduce interface energy. 
As a result, the system will be more thermodynamically stable and melt at a higher temperature $T_{m'}$. 
When the sample cell was cooled to the lowest temperatures, crystals grew and interfacial energy decreased further. 
The crystals were well annealed during cooling under the solid-liquid coexistence condition, resulting in an additional reduction in thermodynamic energy, so that a large hysteresis loop was observed and $T_m$ became higher than the melting temperature $T_{m'}$.

\subsection{Phase Diagram}

\begin{figure}[tb]
\begin{center}
\includegraphics[width=80mm]{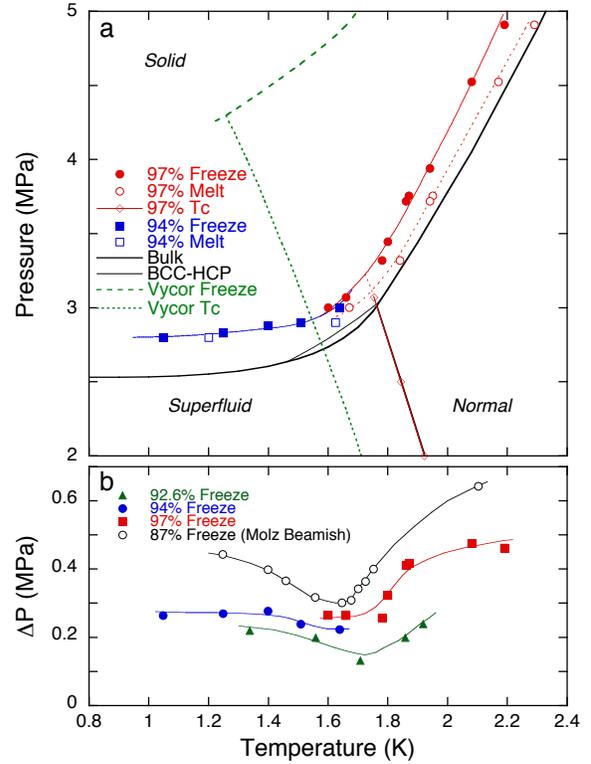}
\end{center}
\caption{(Color online) (a) Phase diagrams of $^4$He in 94.0 and 97.0\% porous aerogels. 
Solid symbols show the phase boundary of freezing, open symbols show melting. 
The superfluid transition temperature is also shown for the 97.0\% porous aerogel. 
The phase boundary in the 92.6\% porous aerogel is close to that of the other aerogels, so, for clarity, data is only shown for two aerogels. 
The superfluid transition temperature is only shown for the 97.0\% porous aerogel, because the lines for the three other aerogels overlap with each other. 
The freezing phase boundary and superfluid transition in Vycor glass are included for comparison. 
(b) Pressure elevation for freezing, with values for the 87\% porous aerogel\cite{MolzBeamish} also plotted. 
The solid lines are the guide for the eye. 
}
\label{PhaseDiadP}
\end{figure}

The melting and freezing points of the three aerogels studied in this work are very similar. 
Figure~\ref{PhaseDiadP}(a) depicts the obtained phase diagrams of $^4$He in the 94.0 and 97.0\% porous aerogels, along with the phase boundary of bulk $^4$He and $^4$He in Vycor glass\cite{ZhaoReppy}. 
The solid symbols mark the onset of freezing and the open symbols represent the completion of melting. 
The freezing curves rapidly become flat at around 1.7~K, making it very difficult to obtain the phase boundary at low temperatures, because small changes in sample cell pressure cause no freezing. 
Bittner and Adams have shown that helium does not melt until the pressure becomes lower than the freezing pressure using a deformable cell, which allows {\it{in situ}} changes in pressure on the sample \cite{BittnerAdams}. 
It is difficult to map out the true melting boundary with our constant-volume cell in the low-temperature region, where the melting curves are relatively flat. 

For the freezing of $^3$He near the minimum melting pressure, it was shown that the temperature of the minimum is not affected in porous materials\cite{BittnerAdams}. 
Then, in helium, it is more appropriate to discuss the effect of confinement in pores with an elevation in $p_f$, rather than with a decrease in $T_f$. 
Figure~\ref{PhaseDiadP}(b) shows the elevation in $p_f$ from the bulk value for the 92.6, 94.0, and 97.0\% porous aerogels, along with the results for the 87\% porous aerogel by Molz and Beamish\cite{MolzBeamish}. 
The pressure elevation in aerogel increased with increasing temperature above 1.7~K, and a similar temperature variation was observed in Vycor glass\cite{BittnerAdams}. 

The homogeneous nucleation theory of droplet formation provides an interpretation of the freezing pressure elevation\cite{ZhaoReppy}. 
The solid droplet can grow, if the decrease in volume energy overcomes the increase in surface energy. 
Solid helium does not wet the pore surface\cite{Balibarinterfacetension}; thus, the size of the nucleation center is limited by pore size. 
If the nucleation occurs at a constant temperature, the additional pressure required $\Delta P$ is expressed as 
\begin{equation}
\Delta P = \frac{2\sigma}{r_c}\left( \frac{v_s}{v_l-v_s} \right) ,
\label{dP} 
\end{equation}
where $r_c$ is the critical radius of the solid, $\sigma$ is the solid-liquid interfacial tension, and $v_s$ and $v_l$ are the solid and liquid molar volumes, respectively. 
In porous materials, freezing occurs in the largest pore, and the pressure elevation is inversely proportional to pore size. 
However, we found no clear porosity dependence in our dilute aerogels. 
It is difficult to detect the difference in pressure elevation between aerogels with large pores, because $\Delta P$ is small in inverse proportion to critical radius. 
It is plausible that there were larger pores for the freezing onset in the 92.6\% porous aerogel because of the wide distribution of pore sizes and differences between the conditions of synthesis. 
Using the pressure elevation and eq.~(\ref{dP}), it is shown that our aerogel had a critical radius between 12 and 15 nm. 
The small pressure elevation in our samples indicates that our aerogels have larger pores than the 87\% porous aerogel\cite{MolzBeamish}, on the basis of the silica density. 
Although pore size distribution is expected to vary with porosity, a sharp freezing onset was observed. 
The nucleated crystal can grow rapidly owing to the open structure of the aerogel.

$T_c$ in 97.0\% porous aerogel is also indicated in Fig.~\ref{PhaseDiadP}(a). 
The sharp critical attenuation peak at the superfluid transition is attributed to the openness of the aerogel structure. 
$T_c$ decreased with decreasing porosity of the aerogel, and shifted by about 10 mK in the 92.6\% porous aerogel\cite{MatsumotoJETP}. 
The pressure dependence of the superfluid suppression was very weak. 

From the perspective of phase transitions in confined geometry, a size effect is expected for both solid-liquid and superfluid transitions. 
A strong correlation exists between $T_c$ and $p_f$ observations in various porous materials\cite{BittnerAdams, Hiroi, MolzBeamish, Pearce, YamamotoShirahama}. 
In comparison with other porous materials, a slight suppression of $T_c$ and freezing pressure elevations in aerogels is elucidated to a considerable extent by pore size. 
When $T_c$ is compared with the data relating transition temperature to pore size reported by Wada et al.\cite{Wada}, the pore sizes estimated for the 92.6 to 97.0\% porous aerogels vary from about 8.5 to 16 nm in radius. 
These pore sizes, estimated from superfluid transition temperature, are in reasonable agreement with those obtained from the elevation in $p_f$, as described above. 

\subsection{Sound transmission through solid-liquid interface} 

As mentioned earlier, sound transmission through the solid-liquid interface in aerogels was observed for the first time in this study. 
The attenuation decreased with decreasing temperature after an abrupt increase at the freezing point. 
Attenuation in the liquid phase was almost the same as that in the solid phase at the lowest temperature at frequencies of 6 and 10 MHz, as shown in Figs.~\ref{970-6MHz_AttC}(a) and \ref{97_10MHz_pure_016_Att}(a). 
Then, the additional attenuation in the solid-liquid coexisting case is caused by the interface, which increased with frequency and seemed to have a weak temperature dependence. 
The experimentally obtained attenuations due to the interface at 0.5~K were about 25~dB at 6~MHz and 28-37~dB at 10~MHz. 
The scattering of the attenuation value results from a variety of interface conditions. 
The interface location changed with solid-to-liquid ratio, and the angle to the direction of sound propagation may vary slightly from one crystal to another. 

Next, we investigated the large sound attenuation at the interface. 
The crystallization of $^4$He in aerogel is thought to affect sound transmission through the interface. 
A large ultrasonic attenuation at the interface has been observed in bulk $^4$He\cite{Castaing, Moelter, Poitrenaud, Amrit} and ascribed to rapid crystallization\cite{Puech,Nozieres}. 

Considering crystal growth at the interface, 
the sound transmission coefficient is given by 
\begin{equation}
\label{transmissioncoeff}
\tau\left(T,\omega\right) = \frac{4 z_{s} z_{l}}{\left\{ (z_{l} + z_{s}) + (z_{l} z_{s}/Z_{i})\right\}^2} . 
\end{equation}
The amplitude of the energy transmitted from one phase to another depends not only on the acoustic impedances of the solid ($z_s$) and liquid ($z_l$), but also on the complex acoustic impedance\cite{Poitrenaud, Amrit} of the interface, $Z_i$, expressed as
\begin{equation}
 Z_i = \left( \frac{\rho_s{\rho_l}^2}{\rho_s - \rho_l}\right) \left\{K^{-1}+\left(im_{I}\omega / \rho_l\right)\right\} , 
\end{equation}
where $K$ is the crystal growth coefficient. 
$\rho_s$ and $\rho_l$ are the densities of the liquid and solid, respectively.
$m_{I}$ is the surface inertia of the crystal.

When crystal growth in the aerogel occurred as rapidly as that in the bulk, the attenuation at the interface was estimated to be as large as 40$\sim$50 dB at 0.5 K using the bulk properties and eq.~(\ref{transmissioncoeff}). 
On the other hand, when crystal growth is completely suppressed in the aerogel, attenuation will be reduced to around 2 dB. 
The experimentally obtained attenuation was smaller than that observed with rapid crystallization in the bulk, but much larger than that expected with suppressed crystallization. 

It has been shown experimentally and theoretically\cite{Castaing, Moelter, Poitrenaud, Amrit, Puech, Nozieres} that growth resistance, the inverse of crystal growth coefficient, of a rough surface can be described by 
\begin{equation}
K^{-1} = A + BT^4 + C\exp{\left(-\Delta/T\right)} ,
\end{equation}
where $\Delta \sim$ 7.2 K is the roton gap. 
The first term A depends on the quality of the crystal, and has been shown to be negligible for good-quality crystals in bulk\cite{Amrit}. 
The second and third terms are caused by phonon and roton scatterings, respectively. 
In the bulk, phonon scattering is dominant below 0.55 K. 
These terms give rise to the temperature dependent crystal growth, as is observed in bulk $^4$He. 
A weak temperature dependence of attenuation indicates that the first term is dominant. 
It is plausible for the crystal in aerogels to have a high degree of disorder due to silica strands. 

Mass transport to the interface is necessary for crystal growth, because of the density difference between solid and liquid. 
Coupling between fluid motion and silica strands was observed below 1 K\cite{MatsumotoJETP} where no normal fluid component substantially exists. 
Reduced mass transport by silica strand would suppress crystal growth, even in the superfluid. 

\subsection{Solid helium in aerogel}

The contribution of solid helium to the total density and elastic modulus as a coupled system of aerogel and solid helium was comparatively large for dilute aerogels. 
Our sample had a large velocity change, and the velocity of solid in the aerogel was close to that of bulk solid. 
The total velocity increase was about 90\% of velocity of liquid in aerogel under high-$p_f$ condition, where $^4$He in the acoustic cavity was completely frozen. 
This increase is much larger than the approximately 23\% increase observed in 87\% porous aerogel\cite{MolzBeamish}. 

Figure~\ref{97_10MHz_pure_016_Att}(a) shows the attenuation of pure $^4$He in the 97.0\% porous aerogel at 10 MHz. 
The attenuation and sound velocity in the other aerogels with various porosity had qualitatively similar temperature variations. 
Attenuation increased with frequency, but the sound velocities at 6 and 10 MHz were the same. 
The temperature dependence of attenuation in the samples was similar to that reported in 87\% aerogel\cite{MolzBeamish} and also to that observed in bulk solid by Goodkind\cite{LenguaGoodkind, HoGoodkind}. 

The attenuation of solid $^4$He in pores is generally ascribed to a stress-relaxation process that involves vacancies in solid\cite{BeamishElbaum,MolzBeamish}. 
The angular frequency $\omega$ dependence of attenuation was explained as a function of relaxation time due to the thermal activation of vacancy motion, and peak attenuation occurs when $1/\omega$ equals the relaxation time. 
Although attenuation decreased with decreasing frequency, as shown in Figs.~\ref{970-6MHz_AttC}(a) and \ref{97_10MHz_pure_016_Att}(a), no attenuation peak was observed.

\subsection{Effect of $^3$He impurity on the interface and in solid $^4$He}

$^3$He impurity in bulk $^4$He has the following effects on the solid and interface. 
$^3$He atoms selectively remain at the solid-liquid interface, limiting crystal growth in the bulk\cite{Wang,MSuzuki}. 
$^3$He atoms in bulk solid $^4$He also have a significant influence on the motion of dislocations\cite{HSuzuki1}. 
In order to study crystal growth and sound attenuation mechanisms in solid, we conducted 10 MHz ultrasound measurements, adding 0.16\% $^3$He impurity. 

\begin{figure}[tb]
\begin{center}
\includegraphics[width=80mm]{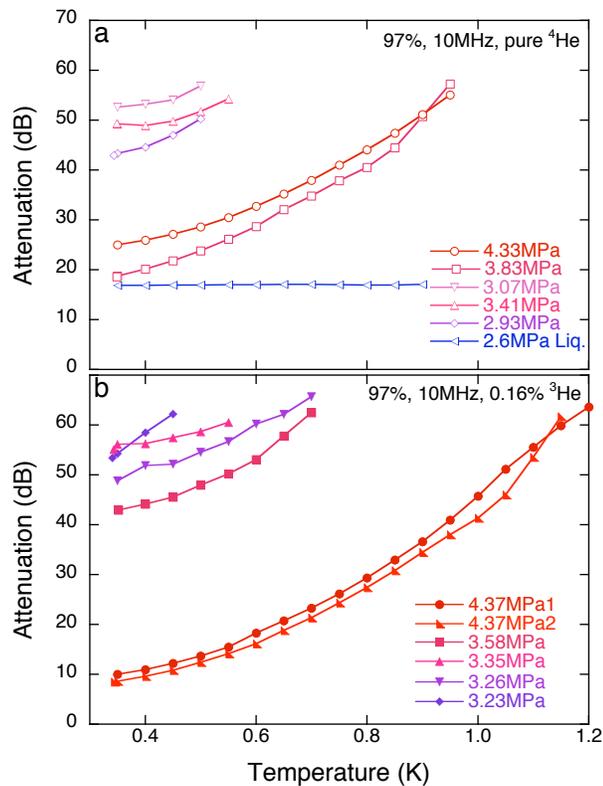}
\end{center}
\caption{(Color online) Attenuations of 10~MHz ultrasound for (a) pure $^4$He and (b) $^4$He with 0.16\% $^3$He in 97.0\% porous aerogel at various freezing pressures $p_f$. 
The attenuation of pure liquid $^4$He at 2.6 MPa is also shown in (a) for comparison. 
}
\label{97_10MHz_pure_016_Att}
\end{figure}

Figure~\ref{97_10MHz_pure_016_Att}(b) shows the attenuation with 0.16\% impurity $^3$He in the 97.0\% aerogel. 
The attenuation at the solid-liquid interface, observed for the range of $p_f$ between 3.23 and 3.58~MPa, was not affected by $^3$He addition. 
This indicates that the crystal growth at the interface has no significant effect on sound transmission. 
It is possible that the disorder at the interface scatters phonons, giving rise to a large attenuation of sound. 
There should be a microscopic disorder at the interface, because solid helium does not wet glass \cite{Balibarinterfacetension}. 

The motion of crystal dislocations induced by an acoustic field is known as an acoustic wave attenuation mechanism in bulk solid $^4$He. 
Iwasa and Suzuki\cite{HSuzuki1} reported that doping about 1\% $^3$He into bulk solid $^4$He reduces sound attenuation, because $^3$He atoms pin the motion of dislocations. 
Helium crystals in aerogels should be highly disordered and should contain many dislocations. 
In our solid samples at $p_{f}=$ 4.37~MPa, $^3$He impurity greatly reduced attenuation, but did not change temperature variation compared with pure $^4$He at 4.33~MPa, as shown in Fig.~\ref{97_10MHz_pure_016_Att}. 
Dislocations in $^4$He in aerogel were pinned by $^3$He atoms as in the bulk. 
The limited variation in temperature dependence indicated that thermally activated vacancies determine the temperature dependence. 

\section{Conclusions}

The velocity and attenuation of longitudinal ultrasound of $^4$He in aerogel were measured to determine  freezing pressure. 
Freezing pressures in several dilute aerogels were elevated by about 0.3~MPa from that of bulk. 
This pressure elevation was considerably smaller than that observed in other porous materials, and almost independent of aerogel porosity above 90\%. 

The transmission of sound through the solid-liquid interface in aerogels has been observed for the first time. 
The ultrasound attenuation at the interface was large, and showed no change with the addition of $^3$He impurity. 
This implies that the sound attenuation at the interface is due to the disorder arising from the aerogel, not from crystal growth.
However, the addition of $^3$He to the solid significantly decreases attenuation. 
The motion of dislocations was pinned by $^3$He, and phonon scattering was suppressed in solid $^4$He; similar mechanisms have been observed in bulk $^4$He. 

Many open questions remain.
It is interesting how a macroscopic interface is generated in aerogel. 
We obtained no detailed information about the crystal growth during cooling, owing to large sound attenuation. 
In order to reduce attenuation, the use of lower frequencies and/or thinner samples may be necessary. 
The aerogel structure, such as fractal dimension, may affect sound propagation in the solid. 
More systematic and precise experiments will reveal the nature of sound propagation, crystal nucleation, and crystal growth, as many porous materials with various pore sizes and structures have recently become available. 

\section*{Acknowledgment}
We thank K. Mukai, T. Tsunekawa and K. Nunomura for experimental assistance. 
This work was supported in part by a Grant-in-Aid for Scientific Research from the Ministry of Education, Culture, Sports, Science, and Technology of Japan.

\end{document}